\documentstyle[preprint,aps,epsfig,amsbsy,floats]{revtex}
 \tightenlines

\begin{document}
\newcommand{\beq}{\begin{equation}}
\newcommand{\eeq}{\end{equation}}
\newcommand{\beqa}{\begin{eqnarray}}
\newcommand{\eeqa}{\end{eqnarray}}
\newcommand{\sr}{\sqrt}
\newcommand{\fr}{\frac}
\newcommand{\mn}{\mu \nu}
\newcommand{\G}{\Gamma}

\draft \preprint{hep-th/0310176,~INJE-TP-03-10}
\title{ Logarithmic corrections to  three-dimensional
black holes and de Sitter spaces}
\author{  Y.S. Myung\footnote{E-mail address:
ysmyung@physics.inje.ac.kr}}
\address{
Relativity Research Center and School of Computer Aided Science, Inje University,
Gimhae 621-749, Korea}
\maketitle

\begin{abstract}We calculate logarithmic corrections to the Bekenstein-Hawking
entropy  for  three-dimensional  BTZ black hole with $J=0$ and
Kerr-de Sitter (KdS) space with $J=0$ including the
Schwarzschild-de Sitter (SdS) solution due to thermal
fluctuations. It is found that there is no distinction between the
event horizon of the BTZ black hole and the cosmological horizon
of KdS space. We obtain the same correction to the Cardy formula
for BTZ, KdS, and SdS cases. We discuss  AdS/CFT and dS/ECFT
correspondences in connection with logarithmic corrections.

\end{abstract}

\newpage
\section{Introduction}
Recently there are several works which show that for a large class
of AdS  black holes, the Bekenstein-Hawking entropy receives
logarithmic corrections due to thermodynamic
fluctuations~\cite{KM1,KM2,CAL,GKS,BS,MS}. The corrected formula
takes the form \beq \label{CEN} S=S_0-\fr{1}{2} \ln \Big(C_v
T^2\Big)+ \cdots, \eeq where $C_v$ is the specific heat of a given
gravitational system at constant volume and $S_0$ denotes the
Bekenstein-Hawking entropy. Here an important point is that for
Eq.(\ref{CEN}) to make sense, $C_v$ should be positive. However,
the d-dimensional (dD) Schwarzschild black hole which is
asymptotically flat has a negative specific heat of
$C_v^{Sch}=-(d-2)S_0$ and thus its canonical ensemble is
unstable~\cite{DMB,GS}. This means that the Schwarzschild black
hole is never in thermal equilibrium with its environment. But
introducing a negative cosmological constant
$\Lambda=-(d-1)(d-2)/2\ell^2$  could make the specific heat
positive and render Eq.~(\ref{CEN}) applicable. The 5D
Schwarzschild-AdS black hole with the event horizon $r^2_{EH}
=\ell^2(-1 + \sqrt{1+4m/\ell^2})/2$ belongs to this
category~\cite{MP,LNOO}. In the limit of large $\Lambda(\ell \to
0,~m>>\ell^2)$, one has a large AdS black hole with $C_v^{SAdS}
\simeq 3S_0$, while in the limit of small $\Lambda(\ell \to
\infty, m<<\ell^2)$, one finds a small AdS black hole which leads
to the Schwarzschild black hole with $C_v^{SAdS} \simeq -3S_0$.
This implies that for a black hole with a large mass, the presence
of a large negative cosmological constant is sufficient to make
the specific heat positive.

In other words, a large black hole in asymptotically AdS space has
a positive specific heat, while a small back hole in
asymptotically AdS space has a negative specific heat. Roughly
speaking, one considers the AdS space as a confining box. Then a
large black hole in a small box (large AdS black hole) can be
thermal equilibrium and  gives a positive specific heat as a
whole. On the contrary, if the box is large and unbounded as in
the small AdS black hole ($\simeq$ Schwarzschild black hole), its
specific heat  becomes negative and thus  this system  could not
be in thermal equilibrium. Similarly one considers  a cosmological
horizon in 5D Schwarzschild-de Sitter space~\cite{NOO}. In the
limit of large $\Lambda(\ell \to 0)$ one finds a positive specific
heat of the pure de Sitter space $C_v^{SdS} \simeq 3 S_0$, while
in the limit of small $\Lambda (\ell \to \infty) $ one recovers a
negative specific heat of the Schwarzschild black hole as
$C_v^{SdS} \simeq -3 S_0$. The cosmological horizon of 5D
hyperbolic topological de Sitter (HTdS) space~\cite{CAI1} has
 the two limits~\cite{myu1,myu2}. Also we expect a similar property in
 topological Reissner-Nordstrom de Sitter space~\cite{SET}.

However, the above two limits  does not exist for a
lower-dimensional gravity system. In other words, this system  is
simple in compared with 5D gravity systems.  A class of 3D
gravitational systems  have  no a singularity but a conical defect
(excess) inside the single horizon. Also A(dS)/(E)CFT
correspondences can be easily confirmed in connection with
logarithmic corrections. In this work, we study 3D AdS black hole
and de Sitter spaces which give positive specific heats and thus
logarithmic corrections to the entropy could be achieved. These
are the BTZ black hole with $J=0$ and Kerr-de Sitter (KdS) space
with $J=0$ including the Schwarzschild-de Sitter (SdS) solution.
All of them have single horizons.

\section{3D AdS black hole}
It is believed that a higher-dimensional black hole in
asymptotically flat spacetime should have a spherical horizon.
When introducing  a negative cosmological constant, a
higher-dimensional black hole could have a non-spherical horizon.
We call this the topological black hole\cite{TADS,BIR}. However,
in three dimensions we could not construct various horizon
geometries because one-dimensional horizon is locally flat and is
equivalent to each other. Instead, we introduce an AdS black hole,
BTZ black hole with $J=0$~\cite{BTZ} as
 \beq ds^{2}_{AdS}=
 -h_{AdS}(r)dt^2 +\fr{1}{h_{AdS}(r)}dr^2 + r^2 d\theta^2,
\label{3BMT} \eeq where  a metric function $h_{AdS}(r)$ is given
by \beq h_{AdS}(r)=-\mu_{AdS}+ \fr{ r^2}{\ell^2} \eeq with a
reduced mass  $\mu_{AdS}= 8 G_3 {\cal M}_{AdS}$ for a massive
point particle located at $r=0$ inside the event horizon. Here we
choose a circle ($S^1$) for the horizon geometry.  The event
horizon is given by
 \beq \label{EH} r_{EH}=
\ell \sqrt{\mu_{AdS}}. \eeq
 Relevant thermodynamic quantities: free energy
($F$), Bekenstein-Hawking entropy ($S_0$), Hawking temperature
($T_H$), specific heat ($ C_v=(dE/dT)_V$) and energy (ADM mass:
$E=M$) are given by \beqa \label{TQ}
&&F=-\fr{\mu_{AdS}}{8G_3},~~ S_0=\fr{\pi \ell \sqrt{\mu_{AdS}}}{2G_3}=C_v,~~\\
\nonumber &&T_H=\fr{\sqrt{\mu_{AdS}}}{2 \pi
\ell}=\Big[\fr{G_3}{\pi^2\ell^2}\Big]S_0,~~E=F+T_H S_0
=\fr{\mu_{AdS}}{8G_3}={\cal M}_{AdS}=M_{AdS}, \eeqa where $G_3$ is
the 3D Newton constant. Here we obtain a positive specific heat.
In this case we have the ADM mass $M_{AdS}={\cal M}_{AdS}$ which
is consistent with that from the Brown-York approach in
Ref.\cite{BK}.

\section{3D Schwarzschild-de Sitter spacetime}
 The
3D Schwarzschild-de Sitter spacetime is given by~\cite{SSV}
 \beq ds^{2}_{SdS}=
 -h_{SdS}(r)dt^2 +\fr{1}{h_{SdS}(r)}dr^2 +r^2 d\theta^2
\label{2BMT} \eeq  where $h_{SdS}(r)$ is given by \beq \label{sdh}
h_{SdS}(r)=1-\mu_{SdS}- \fr{ r^2}{\ell^2} \eeq with a reduced mass
$\mu_{SdS}=8G_3{\cal M}_{SdS}$. The cosmological horizon is
located at
 \beq \label{CH} r_{CH}=
\ell \sqrt{1-\mu_{SdS}}. \eeq In this case  we have to put a
particle with a small mass at $r=0$
($r_{EH}<\ell,~0<\mu_{SdS}<1$). This describes a conical defect
spacetime with a deficit angle $ 2 \pi(1-\sqrt{1-\mu_{SdS}})$,
indicating a world with a positive cosmological constant and a
pointlike object with mass ${\cal M}_{SdS}$~\cite{BMM}.  In the
case of $\mu_{SdS}>1$, any cosmological horizon does not exist.

Interesting thermodynamic quantities for the cosmological horizon
are given by~\cite{myu3} \beqa \label{2TQ}
&&F=-\Big[\fr{1-\mu_{SdS}}{8G_3}\Big],~
S_0=\fr{\pi \ell \sqrt{1-\mu_{SdS}}}{2G_3}=C_v,~\\
\nonumber &&T_H=\fr{\sqrt{1-\mu_{SdS}}}{2 \pi
\ell}=\Big[\fr{G_3}{\pi^2\ell^2}\Big]S_0,~E=F+T_H S_0
=\fr{1-\mu_{SdS}}{8G_3}=M_{SdS}. \eeqa  We obtain a positive
specific heat only for a small massive particle ($\mu<1$) inside
the cosmological horizon. In other words, this gravitating system
composed of the cosmological horizon $r_{CH}<\ell$ and a massive
particle with mass ${\cal M}_{SdS}$ is thermodynamically stable
because $F<0$ when $\mu<1$. If ${\cal M}_{SdS}=0$, we recover an
ADM mass of $M_{dS}=1/8G_3$ for the pure de Sitter space with a
cosmological horizon at $r_{CH}= \ell$. In the case of ${\cal
M}_{SdS}\not=0$, the ADM mass ($M_{SdS}=1/8G_3-{\cal M}_{SdS}$) of
this system is less than that of the pure de Sitter space. Here we
address why a particle with ${\cal M}_{SdS}>0$ inside the
cosmological constant contributes to $M_{SdS}$ as a negative one.
The reason is that even if the matter of a pointlike object has a
positive energy, the binding energy to the gravitational de Sitter
background can make it negative. That is, a conical defect at
$r=0$ swallows up a part of the spacetime and thus appears to
reduce the net amount of the positive energy stored in the
cosmological constant. In this sense we no longer take ${\cal
M}_{SdS}$ as a true mass (energy) in  SdS space. The true mass is
given by the ADM mass $M_{SdS}$.  As a result, the presence of a
pointlike object inside the cosmological horizon decreases the
size of the cosmological horizon in compared with the pure de
Sitter case.

\section{3D Kerr-de Sitter space}
 When
introducing a negative mass in higher-dimensional de Sitter space
one could  find  a non-spherical horizon. We call this the
topological de Sitter space\cite{CMZ}. However, in three
dimensions we could not construct various horizon geometries
because one-dimensional horizon is locally flat and is equivalent
to each other. Here we introduce the Kerr-de Sitter space with
$J=0$~\cite{Park}
 \beq ds^{2}_{KdS}=
 -h_{KdS}(r)dt^2 +\fr{1}{h_{KdS}(r)}dr^2 +r^2 d\theta^2,
\label{BMT1} \eeq where $h_{KdS}(r)$ is given by \beq
h_{KdS}(r)=-\tilde{\mu}_{KdS}- \fr{ r^2}{\ell^2}={\mu}_{KdS}- \fr{
r^2}{\ell^2} \eeq with a negative reduced mass $\tilde \mu_{KdS}
\equiv -\mu_{KdS}= -8 G_3{\cal M}_{KdS}$ in compared with the SdS
case in Eq. (\ref{sdh}). The cosmological horizon is given by
 \beq \label{TCH} r_{CH}=
\ell \sqrt{\mu_{KdS}}. \eeq

Thermodynamic quantities are calculated as \beqa \label{3TQ}
&&F=-\fr{\mu_{KdS}}{8G_3},~ S_0=\fr{\pi \ell \sqrt{\mu_{KdS}}}{2G_3}=C_v,\\
\nonumber &&T_H=\fr{\sqrt{\mu_{KdS}}}{2 \pi
\ell}=\Big[\fr{G_3}{\pi^2\ell^2}\Big]S_0,~E=F+T_H S_0
=\fr{\mu_{KdS}}{8G_3}={\cal M}_{KdS}=M_{KdS}. \eeqa  Here we
obtain a positive specific heat, which means that the KdS space is
thermodynamically stable. We observe  that there is no difference
between Eqs.~(\ref{TQ}) and (\ref{3TQ}).

 Let us ask of how to understand the positivity of  ADM mass
($M_{KdS}>0$) in the presence of a pointlike object with a
negative energy. The topological de Sitter space  in
five-dimensions has a positive mass and a timelike singularity
which remains within a single cosmological horizon. Hence it
satisfies the mass bound conjecture: any asymptotically de Sitter
space whose mass exceeds that of pure de Sitter space contains a
cosmological singularity~\cite{BMM}. In the case of 3D KdS space,
there is no cosmological singularity. However, the negative energy
of a pointlike object at $r=0$ gives the ADM mass of $M_{KdS}=
{\cal M}_{KdS}$ in asymptotically de Sitter space. As a result,
one finds that
$r_{CH}^{KdS}=\ell\sqrt{\mu_{KdS}}>r_{CH}^{dS}=\ell$ if
$\mu_{KdS}>1$. This corresponds to a conical excess. If
$\mu_{KdS}<1$, one finds a conical defect.
%\begin{center}
\begin{table}
 \caption{Summary of specific heat and entropy
 for BTZ black hole, KdS and SdS spaces. Ground-state horizon information ($\mu=0$) is added.}
 %\begin{ruledtabular}
 \begin{tabular}{lp{4.5cm}p{3cm}}
 3D thermal system   & $C_v(=S_0)$ & horizon if $\mu=0(\mu \not=0)$ \\ \hline
BTZ with $J=0$ & + & N \\
KdS with $J=0$ & +  & N \\
SdS & + if $\mu<1$ & Y($\leftarrow$)
 \end{tabular}
 %\end{ruledtabular}
 \end{table}
 %\end{center}
\section{correction to Bekenstein-Hawking entropy}
We summarize our result in TABLE I. In this section we make
corrections to the Bekenstein-Hawking entropy according to the
formula of Eq.(\ref{CEN}). For all cases, one finds $C_v=S_0,~
T_H=\Big[\fr{G_3}{\pi^2\ell^2}\Big]S_0$ without any approximation.
Here we require    a small SdS solution for obtaining a positive
specific heat. All logarithmic corrections to the
Bekenstein-Hawking entropy are given by a single relation as \beq
\label{CENT} S=S_0-\fr{3}{2} \ln S_0 + \cdots. \eeq  This means
that any thermodynamically stable system gets a logarithmic
correction to the Bekenstein-Hawking entropy due to thermal
fluctuation around the equilibrium state. Here we confirm that
this correction is universal for all 3D gravitational systems with
$J=0$. Further we wish to mention that  all thermal quantities of
the KdS spaces take the same form as those from the BTZ black
holes. This implies that there is no distinction between the event
horizon of BTZ black hole and the cosmological horizon of Kerr-de
Sitter space. That is, a pointlike object with a positive energy
inside the event horizon is equivalent to a pointlike object with
a negative energy inside the cosmological horizon. We note that
even for $\mu_{SdS}=0$, the SdS space has a cosmological horizon.
If $\mu_{SdS} \not=0$, the size of the cosmological horizon is
decreased.

\section{AdS/CFT and dS/ECFT correspondence}
 The holographic principle means that the number of degrees of freedom
associated with the bulk gravitational dynamics is determined by
its boundary spacetime without gravity. Let us first discuss the
AdS black hole. The AdS/CFT correspondence represents a
realization of this principle~\cite{HOL}. Brown and Henneaux have
shown that  asymptotic symmetries  in 3D AdS spacetime are
described by a pair of Virasoro algebras with central charges
$c=\bar c =3\ell/2G_3$~\cite{BH}. The generators of Brown-Henneaux
Virasoro algebras are simply  Hamiltonian and momentum constraints
of 3D gravity smeared against appropriate vector fields. For  BTZ
black hole with $J=0$, these are $L_0=M\ell/2=\bar{L}_0$ with the
ADM mass $M={\cal M}_{AdS}$~\cite{CAL}.   In order to calculate
the entropy on the boundary spacetime, we use the Cardy formula
for a pair to yield \beq \label{Cardy} S_{CFT}= 2 \pi \sqrt{\fr{c
L_0}{6}} +2 \pi \sqrt{\fr{\bar{c} \bar{L}_0}{6}}=\fr{ \pi
r_{EH}}{2G_3}=S_0 \eeq which establishes the AdS/CFT
correspondence for the entropy. According to the Carlip's
work~\cite{CAL}, the density of states can be approximated as \beq
\label{den} \rho(L_0,\bar{L}_0)= \fr{8G_3\ell}{r^3_{EH}} \exp
\Big[\fr{ \pi r_{EH}}{2G_3}\Big]. \eeq Thus the entropy correction
due to thermal fluctuation of the  CFT system leads to \beq
\label{Cent} S=S_0
 -3 \ln S_0 + \cdots. \eeq
 Comparing the above with the bulk correction in Eq. (\ref{CENT}), one finds logarithmic
 term that differs from it by a factor of 2. However, this
 difference does not seem to be serious because  other approaches suggest a correct
 factor of 3/2~\cite{DMB,CAL}.

 For  SdS and KdS spaces, according to Strominger~\cite{STR}, the dS/ECFT
 proposal is useful for calculating the logarithmic corrections
 to the boundary entropy. We note that 3D de Sitter space can
 be represented by the group manifold of
 SL(2,C)/SL(2,R). The asymptotic symmetry group of de
 Sitter space, subject to the boundary conditions,
  is the two-dimensional Euclidean conformal group
 SO(3,1), which contains the isometry group SL(2,C) as a subgroup.
 Indeed, the Brown-Henneaux analysis can be formally continued to
 arrive at Virasoro algebras with central charges
$c=\bar c =3\ell/2G_3$~\cite{BMM}. Its dual theory might be a
 Lorentzian CFT with  Euclidean signature Virasoro algebras.
 Actually the ADM masses $M_{SdS},M_{KdS}$ in Eqs.(\ref{2TQ}) and (\ref{3TQ})
 are consistent with those computed with the Euclidean signature on the  spacelike boundary.
 Thus we have  generators as $L_0=M\ell/2=\bar{L}_0$
 where  the ADM mass is given by $M_{SdS}=1/8G_3-{\cal M}_{SdS}$ for
 the SdS case and  $M_{KdS}={\cal M}_{KdS}$ for the KdS space.
 At this stage, we remind the reader that
 for an observer near infinity, there is no actual difference  between the event
 horizon of  AdS black hole and the cosmological horizon of
  SdS and KdS space. The difference results in the precise value of a conserved
 quantity, the ADM mass.
 Hence we expect that their dual ECFTs
 exist on the  boundary.  The Cardy formula provides
 \beq \label{2Cardy} S_{ECFT}=
2 \pi \sqrt{\fr{c L_0}{6}} +2 \pi \sqrt{\fr{\bar{c}
\bar{L}_0}{6}}=\fr{ \pi r_{CH}}{2G_3}=S_0 \eeq which establishes
the dS/ECFT correspondence for the entropy. Following the dS/ECFT
correspondence, their logarithmic corrections will be given by the
same form as in Eq. (\ref{Cent}).

On the other hand, a complete Cardy formula for the asymptotic
density of states of a unitary (E)CFT is given by \beq
\label{CCardy} S_{CCFT}= 2 \pi
\sqrt{\fr{c}{6}\Big(L_0-\fr{c}{24}\Big)} +2 \pi \sqrt{\fr{\bar{c}
}{6}\Big(\bar{L}_0-\fr{\bar c}{24}\Big)}. \eeq A naive application
of this formula with the same central charge and generators fails
to find the Bekenstein-Hawking entropy $S_0$, in contrast with the
use of the Cardy-formula in Eq. (\ref{Cardy}). However, in the
case of $M_{AdS/KdS}\ell/2 >> c/24=\ell/16G_3$,  the complete
Cardy formula is given approximately \beq \label{aCardy}
S_{CCFT}\simeq \fr{\pi \ell \sqrt{\mu_{AdS/KdS}}}{2G_3}={S}_0.
\eeq Hence we find that the Cardy formulae in Eqs.~(\ref{Cardy})
and (\ref{2Cardy})  are considered as  approximate forms to the
complete Cardy formula in Eq.~(\ref{CCardy}).

\section{Discussion}
First of all, we summarize our main result. We find that all
thermal quantities of  KdS space take the same form as those from
BTZ black hole. This means that there is no difference between the
event horizon and cosmological horizon in 3D gravity systems with
the single horizon.

We have a few of comments in order. Any ADM mass for a
gravitational system with a single horizon including the BTZ black
hole with $J=0$, the Kerr-de Sitter space with $J=0$,  and SdS
solution is always positive. In three dimensions, any system with
a single horizon has a positive specific heat and its canonical
ensemble is thermodynamically stable. Furthermore, this system has
an equality of $C_v=S_0$, which dictates that it has a single
horizon in three dimensions. Thus we find the same form for
logarithmic corrections to the Bekenstein-Hawking entropy and  the
Cardy formula.

 Concerning  A(dS)/(E)CFT correspondences,  we note that
 these  are applicable for the event horizon for AdS
 black hole and the cosmological horizons for KdS and SdS spaces.
 Boundary
 systems have the same central charge of $c=\bar c=3
\ell/2G $ with  generators $L_0=\bar{L}_0=M\ell/2$ where $M$ is
one ADM mass for AdS black holes, KdS and SdS spaces. In general
the boundary entropy is equal to the Bekenstein-Hawking entropy
and its logarithmic correction is given by Eq. (\ref{Cent}).

Finally we wish to remark the difference between  3D and 5D
gravity systems. In 5D gravity, the parameter $k$ is introduced to
classify the 3D horizon geometry: $k=0,~1,~-1$ denote the flat,
elliptic, and hyperbolic spaces with constant curvature. That is,
it is necessary to classify  topological AdS black holes and
topological de Sitter spaces. In 3D gravity, however, the horizon
is one-dimensional space and it is locally flat. Hence the
parameter like $k$ is irrelevant to 3D gravity systems. Even if
$k$ is introduced, it could be absorbed into $\mu$ by the
redefinition. Therefore, in this work we investigated thermal
properties of three known examples: BTZ black hole with $J=0$,
Kerr-de Sitter space with $J=0$, and Schwarzschild-de Sitter
space. In other words, the BTZ black hole is representative of 3D
topological AdS black holes, while the Kerr-de Sitter is
representative of 3D topological Sitter spaces.

\section*{Acknowledgments}
The author thanks Rong-Gen Cai and Mu-In Park for helpful
discussions.  This work was supported in part by KOSEF, Project
No. R02-2002-000-00028-0.

\end{document}